\documentclass[aps,prc,twocolumn,superscriptaddress,showpacs,nofootinbib,floatfix]{revtex4}

\usepackage{graphicx}
\usepackage{dcolumn}
\usepackage{bm}

\newcommand{\beq}{\begin{equation}}

\newcommand{\eeq}[1]{\label{#1} \end{equation}}

\newcommand{\beqar}{\begin{eqnarray}}

\newcommand{\eeqar}[1]{\label{#1} \end{eqnarray}}

\newcommand{\bmath}{\begin{displaymath}}

\newcommand{\emath}{\end{displaymath}}

\newcommand{\bitem}{\begin{itemize}}

\newcommand{\eitem}{\end{itemize}}

\begin{document}

\title{\Large \bf Transient field fluctuations effects in d+Au and Au+Au collisions at $\sqrt{s_{NN}}$ = 200 GeV.}

\newcommand{\mcgill}{McGill, University, Montreal, Canada, H3A 2T8}

\newcommand{\columb}{Physics Department,
Columbia University, New York, N.Y. 10027}

\newcommand{\frank}{Frankfurt Institute for Advanced Studies,
J.W. Goethe Universit\"{a}t, D60438 Frankfurt am Main, Germany}

\newcommand{\wayne}{Wayne State University, Detroit, MI 48201, USA}

\affiliation{\mcgill}
\affiliation{\columb}
\affiliation{\frank}
\affiliation{\wayne}

\author{~V.~Topor~Pop} \affiliation{\mcgill}
\author{~M.~Gyulassy} \affiliation{\columb} \affiliation{\frank}
\author{~J.~Barrette} \affiliation{\mcgill}
\author{~C.~Gale} \affiliation{\mcgill}
\author{~S.~Jeon} \affiliation{\mcgill}
\author{~R.~Bellwied} \affiliation{\wayne}

\date{December 11, 2006}


\begin{abstract}

The effect of fluctuations of strong color 
electric fields (SCF) on the baryon production
in {\it d}+Au and Au+Au collisions at 200A GeV  
is studied in the framework of the {\small HIJING/B\=B v2.0} model.
It is shown that the dynamics of the production process deviates
considerably from calculations based on Schwinger-like estimates for 
homogeneous and constant color fields. 
An increase of the string tension from $\kappa_0$ = 1 GeV/fm, 
to {\em in medium mean values of} 1.5 to 2.0 GeV/fm and 2.0 to 3.0 GeV/fm for 
{\it d}+Au and Au+Au, respectively, results in a consistent description of the 
observed nuclear modification factors 
{\it R}$_{{\it d}{\text {Au}}}$ and {\it R}$_{{\text {AuAu}}}$ 
(that relates ({\it d})Au+Au and {\it p}+{\it p} collisions), 
and point to the relevance of fluctuations 
of transient color fields. The differences between  nuclear 
modification factors {\it R}$_{{\text {AuAu}}}$ and {\it R}$_{{\text {CP}}}$ 
(that relates central and peripheral collisions) are also discussed.
The measurement of multi-strange (anti)hyperons ($\Xi$, $\Omega$) 
yields would provide a crucial test of the importance of 
SCF fluctuations at RHIC energies.

\end{abstract}


\maketitle



\newpage

\section{Introduction}

While the phase transition from hadronic degrees of freedom to
partonic degrees of freedom (quarks and
gluons) in ultra-relativistic nuclear collisions is a central focus of
recent experiments at  the Relativistic Heavy Ion Collider (RHIC),
data on baryon and hyperon production has revealed interesting and unexpected 
features that may be of novel dynamical origin.
As an example, a {\em baryon/meson anomaly} 
\cite{Adcox:2001mf,Adler:2003kg,Vitev:2001zn,Gyulassy:2003mc} 
is observed as a large enhancement of the baryon to meson ratio
and as a large difference of the nuclear modification
factor (NMF) between total charged \cite{Adcox:2001jp}  
and neutral pions ($\pi^0$) \cite{David:2001gk} 
at moderate (intermediate) transverse momenta ($2<p_T<6$ GeV/{\it c}).

In previous papers \cite{prc70_top04,prc72_top05} we studied 
the possible role of topological baryon junctions,
\cite{prc70_top04,Kharzeev:1996sq},
and the effects of strong color field (SCF) \cite{prc72_top05} 
in nucleus-nucleus collisions. 
We have shown, in the framework of {\small HIJING/B\=B v2.0} model,
that junction-antijunction ($J\bar{J}$) loops 
with an enhanced {\em intrinsic
transverse momentum}  $k_T \approx $ 1 GeV/{\it c},
a default string tension $\kappa_{0}$ = 1 GeV/fm,
and a diquark suppression factor ($\gamma_{\text {qq}}$=0.07)
provide a partial explanation of the baryon/meson 
anomaly \cite{prc70_top04}.
That model provides an alternative dynamical 
explanation of the data to recombination models \cite{muller03}.  
Within  {\small HIJING/B\=B v2.0} \cite{prc70_top04,prc72_top05} 
one of the main assumptions is that 
strings could survive and fragment \cite{miklos_zf_91,biro_91}, 
and in particular populate the mid to low $p_T$ range.
In contrast, in the recombination picture \cite{muller03} or 
in the hydrodynamical approach \cite{heinz_03}
all coherent strings are assumed to become rapidly incoherent
resulting in a rapid thermalization.

In nucleus-nucleus collisions the color 
charge excitations may be considerably
greater than in nucleon-nucleon collisions due to the almost
simultaneous interaction of several participating nucleons in a row
\cite{miklos_zf_91,Biro:1984cf} and could 
be important even in the case of few binary collisions.
Molecular dynamics models have been used to study the effects of color ropes
as an  effective description of the
non-perturbative, soft gluonic part of QCD \cite{soff_jpg04}.

Recently, the effects of gluon field generated in the wake of hard processes
and through primordial fluctuations of the color charges in the nuclei
have been investigated\cite{kapusta05_1,lappi06_1,asakawa06_1}. 
The physical situation immediately after the
collision bears close analogy to string models of high energy collisions.
It was shown that the initial electric and magnetic fields produced in
high energy hadronic collisions are longitudinal and leads to a novel
string-like description of the collisions and a large topological 
charge density after the collisions \cite{lappi06_1}.
In our model, {\small HIJING/B\=B v2.0},  
there are longitudinal electric fields
induced by a collision, which subsequently decay by quantum pair 
production. The effects of longitudinal magnetic fields are taken 
into account by a specific (like {\em Mercedes star}) 
topology for junction anti-junction $J\bar{J}$ loops \cite{ripka_03}.
Strangeness enhancement
\cite{rafelski_82,rene_04,caines_jpg05,cgreiner_02,armesto96,antai97}, 
strong baryon transport \cite{prc72_top05,csernai01}, 
and increase of intrinsic $k_T$ \cite{soff_jpg04} 
are all expected consequences of SCF.
These are modeled  in our microscopic models as
an increase of the effective  string tension that controls the
quark-anti-quark
 ({\it q}$\bar{\it q}$) and 
diquark - anti-diquark (qq$\overline{\text {qq}}$) pair creation rates
and the strangeness suppression factors \cite{Biro:1984cf}.

In previous studies \cite{prc70_top04,prc72_top05} 
we have focused on longitudinal rapidity distributions  and transverse mass 
(or momentum) spectra of hadrons and hyperons in Au+Au collisions 
at RHIC energies. 
However, the benchmark test of microscopic models are 
{\it p}+{\it p} and {\it p}({\it d})+{\it A} data. 
The results from {\it p}+{\it p} collisions are used as a base line
reference to obtain nuclear modification factor {\it R}$_{\text {AA}}$. 
Therefore, we explore further dynamical effects associated with
long range coherent fields (i.e strong color fields, SCF),
including baryon junctions \cite{Kharzeev:1996sq} and loops 
\cite{prc72_top05,Vance:1999pr,svance99},
in the framework of {\small HIJING/B\=B v2.0} \cite{hij_top06}, with emphasis 
on the novel baryon and hyperon observables measured at RHIC
in {\it p}+{\it p} and {\it d}+Au collisions.
Using this model we analyze baryon/meson anomaly, 
the particle species dependence, centrality dependence 
in {\it d}+Au and Au+Au collisions, as well as the differences seen  
in nuclear modification factors {\it R}$_{\text {AA}}$ (that relates Au+Au and 
{\it p}+{\it p} collisions) 
and {\it R}$_{\text {CP}}$ (comparing central and peripheral collisions).
Comparison of NMF in different collision systems 
should provide information on the hadronization mechanisms.

The paper is organized as follows. In Sec. II 
we review the {\small HIJING/B\=B v2.0} (with SCF) model. 
For clarity, we repeat here 
some of the basics out-line that has been already presented in 
Refs.~\cite{prc70_top04,prc72_top05}. In Sec. III we discuss
theoretical predictions in comparison with recent RHIC 
experimental data. Finally summary and conclusions are given in 
Sec. IV. 
 


\section{HIJING/B\=B v2.0 model}

\subsection{Junction anti-Junction Loops}

{\small HIJING} is a model that provides a theoretical framework
to extrapolate elementary proton-proton multiparticle phenomena 
to complex nuclear collisions
as well as to explore possible new physics such as energy loss and gluon 
shadowing \cite{wang_97}. 
Our analysis are performed in the framework 
of the {\small HIJING/B\=B v2.0} model 
that is based on {\small HIJING/B\=B v1.10} \cite{Vance:1999pr}, 
\cite{svance99} and {\small HIJING} \cite{wang_97}.

In {\small HIJING} \cite{wang_97} 
the soft beam jet fragmentation is modeled by diquark-quark strings 
as in \cite{pyt_94} with gluon 
kinks induced by soft gluon radiations. 
The mini-jet physics 
is computed via an eikonal multiple collision framework
using pQCD {\small PYTHIA 7.3} to compute the initial and final 
state radiation and hard scattering rates. 
In {\small PYTHIA} the cross section for hard parton 
scatterings is enhanced by
a factor {\it K} = 2 in order to simulate high
order corrections. {\small HIJING} extends {\small PYTHIA} to 
include a number of new nuclear effects.
Besides the Glauber nuclear eikonal extension,
shadowing of nuclear parton distributions
is modeled. In addition  dynamical energy loss
of the (mini)jets is taken into account through an effective energy loss, 
$dE/dx$ \cite{wang_97,prc68_top03}.

In {\small HIJING/B\=B v1.10} \cite{Vance:1999pr}
the baryon junction mechanism was introduced as an extension
of {\small HIJING/B} \cite{svance98} in order to try to account
for the observed longitudinal distributions of 
baryons({\it B}) and anti-baryons($\bar{B}$) 
in proton nucleus ({\it p}+{\it A}) and 
nucleus-nucleus ({\it A}+{\it A}) collisions at the SPS energies.
However, as implemented in {\small HIJING/B\=B v1.10} the junction loops
fails to account for the observed enhanced transverse slope of anti-baryons 
 spectra at moderate $p_T$ in {\it A}+{\it A} \cite{prc68_top03}. 
This is due to limitations of the 
$p_T$ algorithm adopted in version 1.10 that includes 
kinematic constraints that worked to oppose the predicted 
enhancement in the baryon junction loop \cite{Kharzeev:1996sq}.
In {\small HIJING/B\=B v2.0} we replaced that algorithm with one that
implements $J\bar{J}$ loops $p_T$ enhancement directly. 
This is done by specifying the intrinsic (anti)diquark
$p_T$ kick in any standard diquark-quark string (qq-{\it q}) that 
contains one or multiple $J\bar{J}$ loops. 
In addition, we introduced 
a new formula (see Eq.(1) below) for generating 
the probability that a given diquark or anti-diquark 
gets an {\em ``enhanced $p_T$ kick''} from the underlying junction
mechanism. We emphasize that while there is very strong 
evidence from a variety
of data that the source of the observed baryon $p_T$ enhancement
arises more naturally from collective hydrodynamic flow \cite{heinz_03},
 elliptic flow of heavy hyperons \cite{castillo_04} argues strongly
for a dominant partonic collective flow as the origin of the baryon anomaly. 
Most of the initial baryon radial $p_T$ could theoretically arise 
from the production mechanism, while its elliptic
deformation would arise from final state interactions.
Nevertheless, our purpose here is to explore more 
fully without the kinematic limitations
of version 1.10, the theoretical problem of 
how much of the baryon anomaly could  be due to
the postulated baryon junction dynamics 
\cite{Kharzeev:1996sq} at RHIC energies and how well is described
({\it p}){\it d}+Au collisions with emphasis on strangeness production.   
The details of this new implementation of $J\bar{J}$ loops 
are described below.

Multiple hard and soft interactions proceed as in {\small HIJING}.
Before fragmentation, we compute via {\small JETSET} \cite{pyt_94} 
the probability that a junction loop occurs in the string.
A picture of a juction loop is as follows: a
color flux line splits at some intermediate point
into two flux lines at one junction and then the flux lines fuse back
into one at a second anti-junction somewhere further 
along the original flux line. 
The distance in rapidity between these points is chosen
via a Regge distribution as described below.
For single inclusive baryon observables this distribution does not need
to be specified.

The probability of such a loop is assumed to
increase with the number of binary interactions, n$_{\text {hits}}$, 
that  the incident baryon suffers in passing through the oncoming nucleus.  
This number depends on the relative and absolute impact parameters 
and is computed in {\small HIJING}
using the eikonal path through a diffuse nuclear density.

We assume that out of the non single diffractive
{\it N}{\it N} interaction cross section, 
$\sigma_{\text {nsd}}$=$\sigma_{\text {in}}-\sigma_{\text {sdf}}$,
a fraction of events $f_{J\bar{J}}=\sigma_{J \bar{J}}/
(\sigma_{\text {in}}-\sigma_{\text {sdf}})$
excite a junction loop (where $\sigma_{\text {sdf}}$ 
is single diffractive cross sections).
The probability after $n_{\text {hits}}$ that the incident baryon
has a $J\bar{J}$ loop is:
\begin{equation}
P_{J\bar{J}}=1-(1-f_{J\bar{J}})^{n_{{\text {hits}}}}
\end{equation}
We take $\sigma_{J \bar{J}}$=17 mb, 
$\sigma_{\text {sdf}} \approx$ 10 mb,
and $\sigma_{\text {in}} \approx$ 42 mb for 
the total inelastic nucleon-nucleon cross section at 
nucleon-nucleon ({\it N}{\it N}) centre of mass (c.m.) energy 
$\sqrt{s_{NN}}$=200 GeV.
These cross sections imply that at $\sqrt{s_{NN}}$=200 GeV, 
a junction loop occurs in {\it p}+{\it p} collisions  
with a rather high probability $17/32\approx 0.5$
and rapidly approaches 1 in {\it A}+{\it A} collisions. 
In {\it p}+{\it S} where $n_{\text {hits}} \approx 2$
there is an 80\% probability that a junction loop occurs in this scheme.
Thus the effects of loops is taken here to have a very rapid onset and
essentially all participant baryons are excited with $J\bar{J}$
 loops in {\it A}+{\it A} collisions at RHIC. 
However, the actual probability is modified 
by string fragmentation due to the threshold cutoff mass
$M_{\text c}= 6$ GeV, which provides sufficient kinematical phase space
for $B\bar{B}$ pair production.

We investigated the sensitivity of the results to the value of parameters   
$J\bar{J}$ and found no significant variation 
on pseudo-rapidity distributions of charged particles produced
in Au+Au collisions at $\sqrt{s_{NN}}$=200 GeV, 
assuming cross sections $\sigma_{J \bar{J}}$,     
15 mb $< \sigma_{J \bar{J}} <$ 25 mb and a cutoff mass $M_{\text c}$,
4 GeV $< M_{\text c} <$ 6 GeV. Light ion reactions such as, 
{\it p}+{\it A} or {\it d}+Au 
should show more sensitivity to $\sigma_{J \bar{J}}$. 
In {\small HIJING/B\=B v2.0}, we introduced \cite{prc70_top04}
the possible topology with two junctions \cite{ripka_03}
and a new algorithm where  
$J\bar{J}$ loops are modeled by an enhancing diquark  $p_T$ kick
characterized by a gaussian  width  
of $\sigma_{\text {qq}}\,'=f \cdot \sigma_{\text {qq}}$, 
where {\it f} is a broadening factor,
and $\sigma_{\text {qq}}$=0.350 GeV/c (consistent with {\small PYTHIA} 
\cite{pyt_94} default value).
In Ref.~\cite{prc70_top04}, we concluded that a value {\it f} = 3 
best reproduced the (anti)proton $p_T$ spectra in Au+Au collisions at 
$\sqrt{s_{NN}}$=200 GeV.
This implementation of the $J\bar{J}$ loops mechanism marks 
a radical departure from that implemented in {\small HIJING/B\=B v1.10}.
While the above procedure allows baryon anti-baryon pairs 
to acquire much higher transverse momentum in agreement with observation,
the absolute production rate also depends on the diquark/quark 
suppression factor $\gamma_{\text {qq}}$ (see Sec. II B). 
The factor {\it f} modifying the default value
$\sigma_{\text {qq}}$=0.350 GeV/c may depend on beam energy, 
atomic mass number ({\it A}), and centrality (impact parameter).
However, we will show that a surprising good description of a variety
of observables is obtained with a constant value, {\it f}=3. 
The sensitivity of the 
theoretical predictions to this parameter is discussed in Sec. III.
Finally, we remark that baryon anti-baryon correlations studies 
in {\it p}+{\it p} and {\it p}({\it d})+Au collisions 
at RHIC energies could eventually help us to obtain
more precise values of $J\bar{J}$ loops parameters 
(mainly: $\sigma_{J\bar{J}}$, Regge intercept $\alpha(0)$, and 
parameter {\it f} ).

Phenomenological descriptions are currently based on
Regge trajectory which gives the appropiate relationship 
between the mass {\it M} of the hadrons and its spin $J_{\text s}$: 
$J_{\text s}=\alpha(0)+\alpha_{\text s}\,\,M^2$,
where $\alpha(0)$ is the Regge intercept, and 
$\alpha_{\text s}$ is the Regge slope.
The value of the Regge slope for baryons   
is  $ \alpha_{\text s}\,\simeq 1$ GeV${^{-2}}$ \cite{svance99} that 
yields a string tension 
(related to the Regge slope, $\kappa_0=1/2 \pi \alpha_{\text s}$)
$\kappa_0 \approx $1.0 GeV/fm.
The multi-gluon exchange processes dominated by Pomeron 
exchange in high energy nucleus-nucleus collisions could be described
by a Regge trajectory with a smaller slope  
$\alpha_{\text s}\,'\, \approx $ 0.45 GeV$^{-2}$ \cite{collins_77}, 
leading to an increase of   
string tension to $\kappa \approx \, 2\,\kappa_0$.

The contribution to the double differential inclusive cross
section for the inclusive production of a baryon and 
an anti-baryon in NN collisions due to $J\bar{J}$ exchange is given by
\cite{Kharzeev:1996sq,Vance:1999pr}: 

\begin{equation}
E_BE_{\bar{B}} \frac{d^6 \sigma } {d^3 p_B d^3 p_{\bar{B}}}
\rightarrow C_{B\bar{B}}e^{(\alpha(0)-1)|y_{B}-y_{\bar{B}}|}
\end{equation}
where $C_{B\bar{B}}$ is an unknown function of the transverse
momentum and  $M_0^J+ P +B$ (junction+Pomeron+baryon) couplings
\cite{Vance:1999pr}.
The predicted rapidity correlation length 
($1-\alpha(0))^{-1}$
depends upon the value of the Regge intercept $\alpha(0)$.
A value of $\alpha(0) \simeq 0.5$ \cite{Kharzeev:1996sq} 
leads to rapidity correlations on the scale $|y_B-y_{\bar{B}}| \sim 2$,
while a value $\alpha(0) \simeq 1.0$ \cite{kopelovic99}
is associated with infinite range rapidity correlations.
Thus, it is important to study rapidity correlations  
at RHIC energies where very high statistics data are now available.

\subsection{Strong Color Field within {\small HIJING/B\=B v2.0}}

In the case of quark-gluon plasma (QGP) creation it is necessary to modify 
the dynamics of particle vacuum production at short time scales 
and the abundance of newly produced particle may deviate considerably 
from the values obtained assuming a constant field 
\cite{zabrodin_04,skokov_prd05}. 
Two possible processes that lead to an increase of 
strangeness production 
within the framework of the Schwinger mechanism are:
i) an increase in the field strength by a modified string tension $\kappa$
\cite{Biro:1984cf,armesto96,soff_99},
or ii) a drop in the quark masses due to chiral symmetry restoration 
\cite{brown_91,bleicher_01,bielich_04,rische_05}.
 A specific chiral symmetry restoration  could be induced by
a rapid deceleration of the colliding nuclei \cite{dima_k05}.
 
For a uniform chromoelectric flux tube with field ({\it E}) the probability 
to create a pair of quarks with mass {\it m}, effective charge 
({\it e}$_{\text {eff}}$),
and transverse momentum ($p_T$) per unit time per unit volume 
is given by \cite{nussinov_80}:
\begin{equation}
P(p_T)\,d^2p_T 
=-\frac{|e_{{\text {eff}}}E|}{4 \pi^3} {\text {ln}} \Bigg\{ 1-{\text {exp}}\left[-\frac{\pi(m^2+p_T^2)}
{|e_{{\text {eff}}}E|} \right] \Bigg\} \,\,d^2p_T
\end{equation}
The integrated probability ($P_m$), when the leading term 
in Eq. 3 is taken into account, is given by:
\begin{equation}
P_m=\frac{(e_{{\text {eff}}}E)^2}{4 \pi^3} \sum_{n=1}^{\infty}
\frac{1}{n^2}{\text {exp}}\left(-\frac{\pi\,m^2n}{|e_{{\text {eff}}}E|}\right)
\end{equation}
where each term in the sum corresponds to production of $n$
coherent pairs, and {\it E} is an homogeneous electric field,
and $\kappa$=$|e_{{\text {eff}}}E|$ is the so called string tension.
We note, that $P_m$ reproduces the classical Schwinger
results \cite{schwinger_51} , derived in spinor quantum
electrodynamics (QED) for describing 
positron - electron ({\it e}$^{+}${\it e}$^{-}$) 
production rate. A sizable rate for spontaneous pair production requires 
``{\em strong electric fields}'', and $|e_{{\text {eff}}}E|/m^2\,\,>$ 1.

Recently, non-perturbative gluon 
and quark-antiquark pair production from a constant 
chromoelectric field with arbitrary color index 
via vacuum polarization have been also investigated in a QCD formalism
\cite{gc_nayak06}.
Although the $p_T$ dependence of the rate of production  
is different because of the presence of the nontrivial 
color generators in QCD, the integration over $p_T$, reproduced also   
Schwinger's result for total production rate used here 
to estimate the suppression of heavier flavors.

The real fields emerging in heavy-ion collisions, could be 
inhomogeneous, with a space-time dependence. Up to now, no reliable
and universal method is available for calculating pair production rates
in inhomogeneous electric fields (for a review see Ref. \cite{gies_prd05}).
Different theoretical methods, such as functional techniques
\cite{fried_prd06} or kinetic equations 
\cite{zabrodin_04,skokov_prd05} have been developed.
It was shown \cite{fried_prd06} that the above  formula 
is a very good approximation 
in the limit of sufficiently intense fields  
if the field does not vary appreciably over space-time distances of less than 
$m^{-1}(m^2/|e_{{\text {eff}}}E|)$. 
In the case of a QGP creation in heavy-ion collisions, 
the characteristic time of the field variation is estimated to be
of order of 1 fm/{\it c} \cite{heinz_03}, and we have to consider 
a time-dependent homogeneous field. 
Taking a soliton-like field, the production probability for strange quarks
is 1.5 times larger than in stationary case \cite{zabrodin_04}, and 
therefore, the abundance of newly produced multi-strange particles 
may considerably deviate from the values obtained for the constant field.

 In {\small HIJING/B\=B v2.0}, as in {\small HIJING},
no attempt has been made thusfar to study
possible modifications of Lund string fragmentation by
back reaction effects as discussed in reference \cite{kluger_back92}.
These effects, important for consideration of final state interaction
kinetic theory, go beyond the scope of the present 
phenomenological investigation.

A value for the strength of the external field, 
based on an estimate of initial energy
density of 50 GeV/fm$^3$ at RHIC  \cite{cooper_plb03}, is 
$E_{\text {ex}} \approx $ 3.16 GeV/fm.
Due to time dependence and fluctuations of the chromolectric field
at the initial stage of the collision we may even expect higher values.
It has been suggested that the magnitude of a typical field strength
at maximum RHIC energies might reach 5--12 GeV/fm \cite{csernai01}.

In general in microscopic string models 
the heavier flavors (and diquarks) are suppressed 
according to Schwinger formula (for homogeneous strong color field, {\it E})
\cite{schwinger_51}:
\begin{equation}
\gamma_{Q}=\frac{P(Q\bar{Q})}{P(q\bar{q})}=
{\text {exp}}\left(-\frac{\pi(m_{Q}^2-m_q^2)}{\kappa}\right)
\end{equation}
where $\kappa=|e_{{\text {eff}}}E|$ is the {\em string tension};
$m_{Q}$ is the effective
 quark mass; {\it Q}={\it s} for strange quark; {\it Q}=qq for a diquark,
and {\it q}={\it u}, {\it d} are the light nonstrange quarks.
 
The main parameters of QCD, the coupling strength $\alpha_{{\text {QCD}}}$ and 
the quark masses, need to be determined precisely. 
However, present estimates \cite{pdb_04} of 
the {\em current quark masses} range from:
$m_{u}$ =1.5--5 MeV; $m_{d}$=3--9 MeV, and $m_{s}$=80--190 MeV. For
diquark we consider $m_{\text {qq}}$=450 MeV \cite{ripka_04_1}.
Taking  for constituent quark masses
of light non-strange quark $M_{u,d}$= 230 MeV,
strange quark $M_{\it s}$=350 MeV \cite{amelin_01},  
and diquark mass $M_{\text {qq}}$=550 $\pm$ 50 MeV 
as in Ref. \cite{ripka_04_1},
it is obvious that the masses of (di)quark and strange quark
will be substantially reduced at the chiral phase transition.
If the QGP is a chirally restored phase of strongly 
interacting matter, 
in this picture, the production of strange hadrons will be enhanced 
\cite{bielich_04}.
In this case, a possible decrease of the strange quark 
mass would lead to a similar enhancement of the suppression factors,
obtained (in microscopic models) by an increase  of string tension 
\cite{zabrodin_04,bleicher_01}.
If the strange quark mass is reduced from $M_{\it s}$=350 MeV to the current 
quark mass of approximately $m_{\it s} \approx \,\,$ 150 MeV
(the actual values are  $m_{\it s}$=80--170 MeV, see Ref.\cite{pdb_04}),
we obtain from Eq. 5 for strangeness suppression factor 
$\gamma_s^1 \,\, \approx\,\,$ 0.70. For a string tension increase
from $\kappa_0$=1.0 GeV/fm to $\kappa$=3.0 GeV/fm, we obtain 
an identical value $\gamma_s^1 \,\, \approx\,\,$ 0.69. 
Moreover, if we consider that Schwinger tunneling could explain
the thermal character of hadron spectra  
and that, due to SCF effects, the string tension value $\kappa$ fluctuates,
we can define an apparent 
temperature as function of the average value of string tension ($ <\kappa>$),
$T=\sqrt{<\kappa>/2\pi}$ \cite{flork_04,raf_06}.

There is a debate in the study of {\it q}{\it q}{\it q} system on the 
relative contribution 
of the $\Delta$-like geometry and the $Y$-like geometry 
\cite{ripka_03,hoft_04},
and on the stability of  
these configurations for the color electric fields \cite{cgreiner_04_y}.
In both topologies we expect a higher string tension than in an
ordinary $q\bar{q}$ string ($\kappa_Y=\sqrt{3}\,\kappa_0$ and 
$\kappa_{\Delta}=(3/2)\,\kappa_0$).
It was shown  \cite{cgreiner_04_y} that the total string tension 
has neither the $Y$ nor the $\Delta$-like value,
but lies rather in-between the two pictures.
However, the {\it Y} configuration appears to be
a better representation of the baryons. 
If two of these quarks stay close together, they behave as a diquark
\cite{hoft_04}.
In a dual superconductor models of color confinement 
for the {\it Y}-geometry, the flux tubes 
converge first toward the centre
of the triangle and there is also another component which 
run in the opposite direction ({\em mercedes star}). 
They attract each other and this 
lower the energy of the {\it Y}-configuration \cite{ripka_03}.

Our calculations are based on the assumption 
that the effective enhanced string tension ($\kappa$), 
in both, basic ropes ($q^n-\bar{q}^n$) and junction ropes
($q^n-q^n-q^n$) are the same. 
For elementary $n$ strings and junctions this ansatz is supported by 
baryon studies \cite{takahashi_05}.  
A different approach to baryon production without baryon junctions
has been proposed in \cite{armesto96} where 
SCF from the string fusion process can lead to 
$(qq)_6-(\bar{q}\bar{q})_{\bar{6}}$ with about a doubling 
of the string tension. 
Both types of SCF configurations may arise but predict different
rapidity dependence of the valence baryons.

Following the discussions above, we 
take into account SCF in our model by an 
{\em in medium effective string tension} 
$\kappa > \kappa_0$, which lead to new values for the suppression factors, 
as well as the new effective intrinsic transverse momentum $k_T$. 
This includes: 
i) the ratio of production rates of  
diquark to quark pairs (diquark suppression factor),  
$\gamma_{\text {qq}} = P({\text {qq}}\overline{\text {qq}})/P(q\bar{q})$,
ii) the ratio of production rates of strange 
to nonstrange quark pairs (strangeness suppression factor), 
$\gamma_{s}=P(s\bar{s})/P(q\bar{q})$,
iii) the extra suppression associated with a diquark containing a
strange quark compared to
the normal suppression of strange quark ($\gamma_s$),
$\gamma_{\text {us}}=(P({\text {us}}
{\overline{\text {us}}})/P({\text {ud}}{\overline{\text {ud}}}))/(\gamma_s)$, 
iv) the suppression of spin 1 diquarks relative to spin 0 ones
(appart from the factor of 3 enhancement of the former based on
counting the number of spin states), $\gamma_{10}$, and 
v) the (anti)quark ($\sigma_{q}''=\sqrt{\kappa/\kappa_0} \cdot \sigma_{q}$)
and  (anti)diquark ($\sigma_{\text {qq}}''= \sqrt{\kappa/\kappa_0} \cdot f
\cdot \sigma_{\text {qq}}$) gaussian  width.
These parameters correspond to $\gamma_{\text {qq}}$=PARJ(1), 
$\gamma_{s}$=PARJ(2), $\gamma_{\text {us}}$=PARJ(3), $\gamma_{10}$=PARJ(4),
and $\sigma_{\text {qq}}$ = $\sigma_{q}$= PARJ(21) 
of the {\small JETSET} subroutines \cite{pyt_94}.
The values of the main parameters sets
 used in our analysis (i.e {\it f}, $\sigma_{q}$,  
as well as  the suppression factors 
$\gamma_{\text {qq}}$, $\gamma_{s}$, $\gamma_{\text {us}}$, and $\gamma_{10}$)
are found in the Appendix.

\section{Numerical results} 

\subsection{Transverse momentum spectrum}
\subsubsection{Identified particles spectra from {\it p}+{\it p} collisions 
at $\sqrt{s_{NN}}$=200 {\rm GeV}}

We will concentrate our discussions on species dependence 
of the nuclear modification factors (NMFs) {\it R}$_{\text {AA}}$ 
and {\it R}$_{\text {CP}}$ in {\it d}+Au and Au+Au collisions 
at $\sqrt{s_{NN}}$=200 GeV.
{\it R}$_{\text {AA}}$ is  the ratio of the heavy-ion yield to 
the {\it pp} cross section normalized by the 
mean number of binary collisions ($<N_{\text {bin}}>$),
while {\it R}$_{\text {CP}}$ is the ratio of scaled central to peripheral 
particle yield. They are defined as in Ref.\cite{prc70_top04}.
By comparing the yields in {\it p}({\it d})+{\it A} and {\it A}+{\it A} 
collisions to that from 
{\it p}+{\it p} collisions, with a scaling factor to take into account the 
nuclear geometry, one can test the assumption 
that nucleus-nucleus collision is a simple superposition of 
incoherent nucleon-nucleon scattering and explore possible nuclear effects
(e.g., shadowing, quenching, SCF).
The relevance of baseline {\it p}+{\it p} hard    
$p_T$ spectra for understanding high-energy
nucleus-nucleus physics is discussed in Ref.~\cite{enteria_05jpg}.
The precision in the first results on the NMFs 
{\it R}$_{\text {AA}}$, were limited by
the uncertainty in the parameterization of the   
{\it p}+{\it p} reference spectrum used in obtaining {\it R}$_{\text {AA}}$.
This limitation was partly overcome when new data on {\it p}+{\it p},
{\it d}+Au and Au+Au collisions 
at $\sqrt{s_{NN}}$=200 GeV using the same experimental setup were obtained
\cite{Adams:2003kv,Adler:2003qi,Adams:2003xp,Adams:2003am,Adams:2003qm,Adams:2006_1,Adams:2006_2,Bellwied:2005bi,Adler:2003au,Adams:2005dq,Adams:2006ke,Adler:2006xd}.

\begin{figure} [h]
\centering
\includegraphics[width=0.8\linewidth]{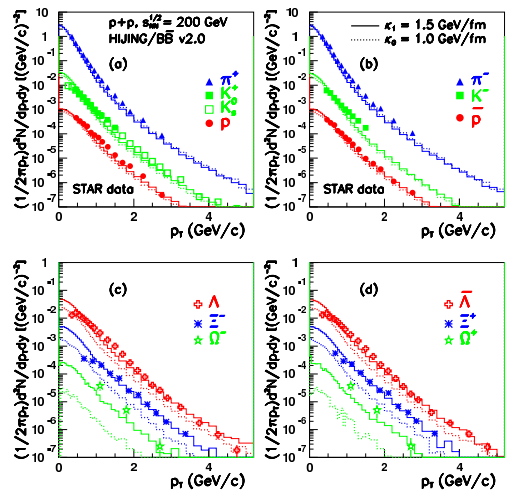}
\caption[p,pi,k pt p+p collisions] {\small 
(Color online) Comparison of {\small HIJING/B\=B v2.0} predictions for 
the invariant yields of pions, kaons (scaled by $10^{-1}$), 
protons (scaled by $10^{-2}$) and 
their anti-particles (upper panels) and (multi)strange 
and their anti-particles (lower panels) as function of $p_T$
from non-single diffractive (NSD) {\it p}+{\it p} colllisions 
at $\sqrt{s_{NN}}$=200 GeV.
The rapidity range was $-0.5 < y < 0.5$.
Results with $\kappa_1$=1.5 GeV/fm (solid histograms; Tab. I, Set 3)
and with  $\kappa_0$=1.0 GeV/fm (dotted histograms; Tab. I, Set 1)
are shown. 
The data are from STAR \cite{Adams:2006_1,Adams:2006_2}. 
Only statistical error bars are shown.
\label{fig:fig1}
}
\end{figure}

Before discussing our results on the NMFs that are scaled yield ratios,
it is important to show how well the model describe {\it p}+{\it p} 
collision data. 
Fig.~\ref{fig:fig1} presents a comparison 
of experimental transverse momentum spectra 
\cite{Adams:2006_1,Adams:2006_2} of positive pions ($\pi^+$), 
kaons ({\it K}$^+$), protons ({\it p}), (multi)strange particles 
 and their anti-particles, with the predictions of 
regular {\small HIJING} (dotted histograms) 
and {\small HIJING/B\=B v2.0} (solid histograms).
{\small HIJING} results describe rather well the data for produced 
``bulk particles'' (i.e. pions, protons, and kaons).
We would have expected that other (multi)strange particles
to also be well described by the standard fragmentation included in 
{\small HIJING}, since the universality of the fragmentation process between 
positron-electron ({\it e}$^{+}$+{\it e}$^{-}$) and 
{\it p}+{\it p} collisions has been confirmed \cite{kniehl_01npb}.
However, regular {\small HIJING} underestimate significantly the spectra
of lambda and cascade particles (Fig.~\ref{fig:fig1} c,d).

{\small HIJING/B\=B v2.0} results include the contribution
of a broadening of the intrinsic momentum ($k_T$),
simulated by an increase of the string tension from the
default value $\kappa_0$=1.0 GeV/fm (Tab.~\ref{tab:tab1}, Set 2) 
to $\kappa_{1}$=1.5 GeV/fm (Tab.~\ref{tab:tab1}, Set 3).
Such parametrization is supported by earlier experimental
measurements \cite{alexopoulos98}, showing that 
at center of mass energy $\sqrt{s_{NN}}\,>$ 100 GeV, 
collisions between protons and anti-protons, 
largely consist of more than a single 
parton-parton interaction. It is also supported  
by a recent PHENIX analysis of the properties of 
jets produced in {\it p}+{\it p} collisions at 
$\sqrt{s_{NN}}\,$=200 GeV \cite{Adler:2006sc},
giving a value of $<k_T>$=$1.34 \pm 0.04 \pm 0.29$ GeV/c,
much larger than the naive expectation  for the pure intrinsic
parton transverse momentum based on nucleon constituent quark mass
(i.e., $<k_T>$ $\approx$ 0.300 GeV/{\it c}) \cite{feynman_77}.
This new parameterization results in an increase of the yield 
of (multi)strange particles without affecting significantly
results for the ``bulk particles'' and thus gives
a simultaneous description of the non-strange and 
multi-strange sector. 


A related approach has been recently presented by 
STAR collaboration \cite{Bellwied:2005bi}.
They introduced an improved description for strange particles
using {\small PYTHIA} by increasing the {\it K} factor 
which quantifies the contributions
of next-to-leading order (NLO) effects, from {\it K=2} to {\it K}=3.
We test this suggestion using our model and observes that such increase of 
{\it K} results in an overestimation of pions, kaons, and protons yields.
In our calculations we keep {\it K} to its default value in 
{\small PYTHIA} and {\small HIJING} i.e., {\it K}=2 for all collisions.


\subsubsection{Identified particles spectra from {\it d}+{\rm Au} 
collisions at $\sqrt{s_{NN}}$=200 {\rm GeV}}

Initial-state nuclear effects are present in both 
{\it d}+Au and Au+Au collisions,
while final state effects are expected to contribute only in Au+Au collisions.
Thus, effects from the initial state are best studied through 
a ``control'' experiment such as {\it d}+Au collisions
\cite{Adler:2003ii,Adams:2003im,Arsene:2003yk,Back:2003ns}.
Multiple soft scattering of projectile partons 
as they traverse a target nucleus may increase  
their transverse momentum 
before they undergo the hard scattering or subsequent to it,
leading  to an enhancement of  the yield at 
moderate and high $p_T$ compared to {\it p}+{\it p} collisions - called 
the Cronin effect \cite{cronin_75,antreasyan_79}.
This enhancement is expected to have some particle mass dependence 
and to be stronger for heavier particles \cite{antreasyan_79}. 
The Cronin enhancement was addressed in Refs. 
\cite{levai_99,wang_00,vitev_02,accardi_04}.
However, these pQCD calculations 
can not predict the particle species dependence
observed in the data, as initial state parton scattering precedes 
fragmentation into the different hadronic species.
An alternative explanation 
considering final state interactions is discussed in Ref.~\cite{hwa_05}.

{\small HIJING} type models incorporate in addition 
to the soft and hard dynamics 
(discussed in Sec. II A) a simulation of soft multiple 
initial state collision effects \cite{miklos_plb98}.
Excited strings are assumed to pick up random transfer 
momentum kicks in each inelastic scattering according to the 
distribution:

\begin{equation}
g(k_c) \propto \{(k_c^2+p_1^2)(k_c^2+p_2^2)
(1+exp[(k_c-p_2)/p_3])\}^{-1}
\end{equation}  

where $k_c$ represents are ntrinsic transverse momentum of the colliding 
partons. The parameters  $p_1$=0.1, $p_2$=1.4, $p_3$=0.4 GeV/c 
are chosen to fit the low energy multiparticle production \cite{wang_97}.
A flag in the code makes it possible to compute 
spectra with and without this effect. 
At RHIC energies  
the observed larger enhancement for protons and anti-protons
in comparison with pions \cite{Adler:2006xd} requires 
new processes beyond the initial state multiple scattering.
In addition to ``Cronin enhancement'' other known initial state effects
that could contribute include nuclear shadowing 
\cite{wang_97,ramona_04,Armesto:2006ph}, 
gluon saturation \cite{dima_03},
$J\bar{J}$ loops, and SCF effects~Ref.~\cite{prc72_top05}.

\begin{figure} [h]
\centering
\includegraphics[width=0.8\linewidth]{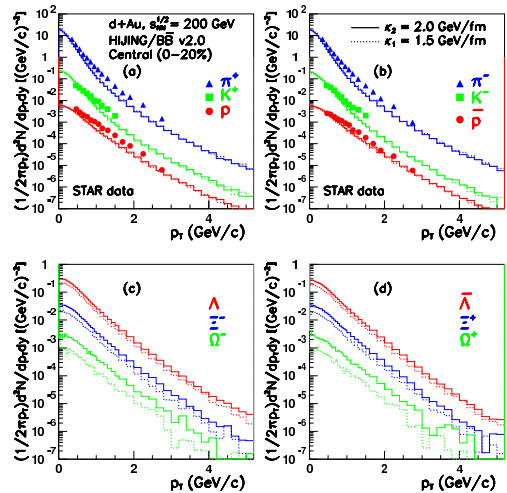}
\vskip 0.5cm
\caption[p,pi.k,strang, d+auc020] {\small (Color online)
Upper part: comparison of {\small HIJING/B\=B v2.0} predictions for 
the invariant yields of pions, kaons (scaled by $10^{-1}$), 
protons (scaled by $10^{-2}$) (left panel) and 
their anti-particles (right panel) as function of $p_T$
for central (0--20\%) {\it d}+Au colllisions at $\sqrt{s_{NN}}$=200 GeV.
Lower part: spectra for strange ($\Lambda$) 
and multi-strange ($\Xi^-$, $\Omega^-$) particles (left panel),
and their anti-particles (right panel). 
The rapidity range was $-0.5 < y < 0.5$.
Results with $\kappa_2$=2.0 GeV/fm (solid histograms; Tab. II, Set 4)
and with  $\kappa_1$=1.5 GeV/fm (dotted histograms; Tab. II, Set 3)
are shown. 
The data are from STAR \cite{Adams:2006_1}.
Only statistical error bars are shown.
\label{fig:fig2}
}
\end{figure}

\begin{figure} [h]
\centering 
\includegraphics[width=0.8\linewidth]{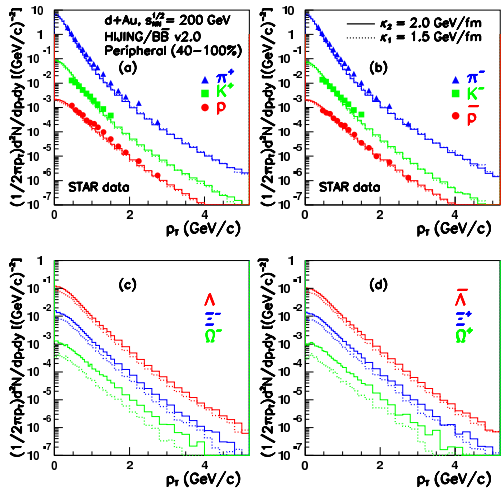}
\caption[p,pi,k ,strand d+Auc40-100] {\small (Color online)
Results for peripheral (40--100\%) d+Au colllisions at $\sqrt{s_{NN}}$=200 GeV.
The histograms have the same meaning as in Fig.~\ref{fig:fig2}.
The data are from STAR \cite{Adams:2006_1}.
Only statistical error bars are shown.
\label{fig:fig3}
}
\end{figure}

In order to further understand the mechanisms responsible for
the particle $p_T$ spectra, and to separate the effects
of initial and final partonic rescatterings we study 
the transverse momentum spectra at mid rapidity   
of identified particles in central (0--20\%) (Fig.~\ref{fig:fig2})
and peripheral (40--100\%) (Fig.~\ref{fig:fig3}) {\it d}+Au collisions.
The data are compared to the predictions of our model assuming 
two values of the string tension,
$\kappa_2$=2.0 GeV/fm (solid histograms) and $\kappa_1$=1.5 GeV/fm
(dotted histograms).
All calculations include nuclear shadowing but do not 
take into consideration ``jet quenching'', since {\it d}+Au collisions
create a ``cold nuclear medium''. 
The agreement above 1 GeV/{\it c}, is very good for 
peripheral ($<N_{\text {bin}}>$=4.5)
and minimum bias ($<N_{\text {bin}}>$=7.8) events (not shown here).
Within our model the observed ``Cronin enhancement'' for 
identified particle spectra
at moderate $p_T$, as well as species dependence, 
are satisfactorily predicted at mid-rapidity
due to a subtle interplay of initial state effects.

However, the calculations somewhat underestimate the pion,
proton and kaon yield in the case of 
central {\it d}+Au collisions ($<N_{\text {bin}}>$=14.8).
To quantify the competing mechanisms 
contributing to the particle yield, we study in the above
results, the contribution from ``soft Cronin'' and shadowing effects.
Including ``soft Cronin'' through Eq. 6 results in an increase of the yield  
which is higher for protons 
($\approx$ 40\% for $p_T$=2--3 GeV/{\it c}) than for pions 
($\approx$ 20\% for $p_T$=2--3 GeV/{\it c}).
On the other hand 
shadowing effects, result in a roughly equal decrease of the yield 
of all species, i.e., $\approx$ 25\% in the same $p_T$ region.
Therefore, these two effects act in opposite directions and partly 
cancel each other at low and moderate $p_T$ ($<$ 5 GeV/{\it c}) . 
In addition, we can not improve the overall description 
of $p_T$ spectra, by modifying 
the parameters from Eq. 6, without destroying the satisfactory results
at low $p_T$ (0.0--1.5 GeV/{\it c}). 
A further increase of mean values of string tension 
to values greater than 2.0 GeV/fm, 
results in a stronger underestimation of the pions yields.  
The description of $p_T$ spectra for central collisions 
point to a mechanism not included in our model.

The disagreement may be explained by considering 
hadronic rescattering in gold nucleus as in 
Multi Phase Transport ({\small AMPT}) model \cite{ampt_lin03},
or inelastic and elastic parton ladder splitting 
as in {\small EPOS} model \cite{Werner:2005jf}.
We do not include here a discussion 
of the particle production at forward rapidity, where possible
gluon saturation effects (color glass condensate CGC) \cite{dima_03}
or parton ladder splitting \cite{Werner:2005jf}, 
may have to be considered.     

In Section III B we will discuss the nuclear modification factor 
for (multi)strange particles. Therefore,
predictions for $\Lambda$, $\Xi^-$, $\Omega^-$ and their antiparticles 
are also shown in Fig.~\ref{fig:fig2} and 
Fig.~\ref{fig:fig3} (lower panels) for the same centrality selections.
We note that in {\it d}+Au collisions the sensitivity to in medium 
string tension depend on mass and strangeness content, 
resulting in a predicted significant  
increase of the yield of multi-strange hyperons ($\Xi$ and $\Omega$).


\subsubsection{Identified particles spectra from {\rm Au+Au} collisions at $\sqrt{s_{NN}}$= 200 {\rm GeV}}

\begin{figure} [h]
\centering
\includegraphics[width=0.8\linewidth]{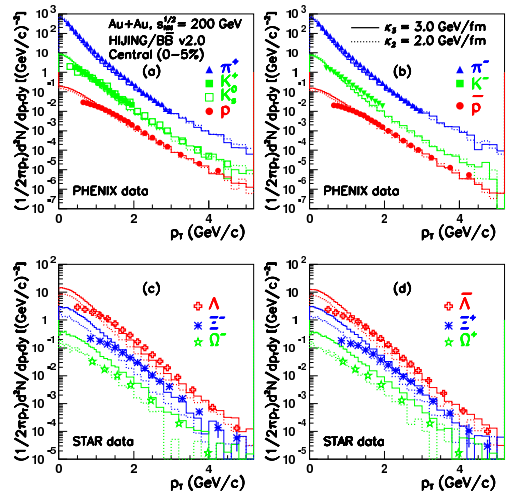}
\vskip 0.5cm
\caption[L] {\small (Color online)
Upper part: comparison of {\small HIJING/B\=B v2.0} predictions for 
the invariant yields of pions, kaons (scaled by $10^{-1}$), 
protons (scaled by $10^{-2}$) (left panel) and 
their anti-particles (right panel) as function of $p_T$
for central (0--5\%) Au+Au colllisions at $\sqrt{s_{NN}}$=200 GeV.
Lower part: spectra for strange ($\Lambda$) 
and multi-strange ($\Xi^-$, $\Omega^-$) particles (left panel),
and their anti-particles (right panel). 
The rapidity range was $-0.5 < y < 0.5$.
Results with $\kappa_3$=3.0 GeV/fm (solid histograms; Tab. II, Set 5)
and with $\kappa_2$=2.0 GeV/fm (dotted histograms; Tab. II, Set 4)
are shown. 
The data are from PHENIX \cite{Adler:2003au} 
and STAR \cite{Adams:2005dq,Adams:2006ke}.
Only statistical error bars are shown.
\label{fig:fig4}
}
\end{figure}

Figure~\ref{fig:fig4} presents a comparison of 
the transverse momentum spectra of identified  
$\pi^+$, {\it K}$^+$, and {\it p} and their anti-particles with 
the predictions of {\small HIJING/B\=B v2.0} (upper panels)
for central (0--5\%) Au+Au collisions.
These data are taken from Ref. \cite{Adler:2003au}.
A similar comparison (lower panels) is also presented for
$\Lambda$s (for central 0--5\%), 
$\Xi^{-}$s, $\Omega^{-}$s (for central 0--10\%) and their anti-particles.
These data are taken from Ref. \cite{Adams:2006ke}.
The shape of the measured spectra show a clear mass dependence.
The protons and anti-protons  
as well as hyperons ($\Xi, \Omega$) spectra have a pronounced shoulder-arm
shape at low $p_T$ characteristic of radial flow.
The results of the model are shown by the dotted ($\kappa_2$=2.0 GeV/fm)
and solid ($\kappa_3$=3.0 GeV/fm) histograms. 
Introducing the new $J\bar{J}$ loops algorithm and SCF effects in
{\small HIJING/B\=B v2.0} results in a significant improvement 
(relative to the predictions without SCF \cite{prc70_top04}) 
in the description of the pion, kaon and proton spectra 
at intermediate $p_T$ for both values of $\kappa$. 
Only a qualitative description is obtained at low $p_T$
due to the presence of elliptic (hydro) and radial 
expansion \cite{heinz_03}, not included in our model.

The yields and transverse momentum slopes of (multi)strange particles 
at moderate $p_T$ are still underestimated 
for the lower value of string tension $\kappa_2$=2.0 GeV/fm.
Multi-strange hyperons yields and spectra, 
seems to favour the larger value $\kappa_3$=3.0 GeV/fm. 
This suggest that multi-strange particles
are produced early in the collisions, when temperature is higher
than those which characterize the ``bulk particles'' 
(i.e., $\pi$, {\it K}, {\it p}). 
This points towards a dynamical origin 
and could be explained as an effect of fluctuations of the  
transient strong color field at early time. Recently, it was also shown that 
``bulk particles'' seems to have a different temperature at kinetic 
freeze-out than hyperons ($\Xi, \Omega$)
suggesting also that multi-strange baryons do not take part 
in ``{\em the same collectivity}'' as  $\pi$, {\it K}, and {\it p}  
during the collision \cite{estienne_05,nuxu_sqm06}. 



\subsection{Nuclear Modification Factors}

\begin{figure} [h]
\centering 
\includegraphics[width=0.8\linewidth]{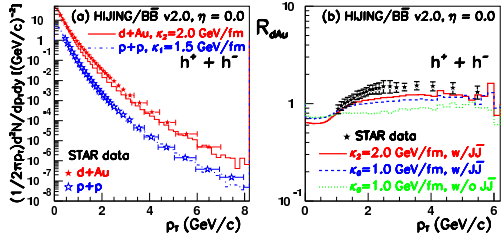}
\caption[p_t,rdAu,h+ + h- ] {\small (Color online)
Left panel: {\small HIJING/B\=B v2.0} predictions with SCF for 
$p_T$ spectra at mid-rapidity of total inclusive charged hadrons,
for central (0--20\%) {\it d}+Au and {\it p}+{\it p} collisions.
The calculations assume $\kappa_2$=2.0 GeV/fm for {\it d}+Au 
(solid histograms; Tab. II, Set 3) and $\kappa_1$=1.5 GeV/fm for 
{\it p}+{\it p} (dot-dashed histograms; Tab. I, Set 3).
Right panel: Nuclear modification factor {\it R}$_{{\it d}{\text {Au}}}$.
The results obtained without $J\bar{J}$ loops and SCF 
(dotted; Tab. II, Set 1),
including only $J\bar{J}$ loops (dashed; Tab. II, Set 2),
and including both effects $J\bar{J}$ loops and SCF 
(solid histograms, Tab. II, Set 3) are shown.
The data are from STAR \cite{Adams:2003xp,Adams:2003im}.
Only statistical error bars are shown.
\label{fig:fig5}
}
\end{figure}

To evaluate if this version of {\small HIJING/B\=B}
describes the produced entropy, the  predicted transverse momentum 
spectra (left panels) and {\it R}$_{{\it d}{\text {Au}}}$ 
(right panels) of the 
total hadron yield for central (0--20\%){\it d}+Au collisions,
are compared to data from Refs.~\cite{Adams:2003xp,Adams:2003im}
in Fig.~\ref{fig:fig5}.
The data could not be well described  
by assuming only a broadening of the {\em intrinsic $k_T$} from 
its standard value $\sigma_{\text {qq}}$=0.360 GeV/{\it c} 
(dotted histograms) to 
$\sigma_{\text {qq}}'$=1.08 GeV/c (i.e., including $J\bar{J}$ loops, 
dashed histograms). The introduction of SCF has a slight 
effect on the predicted nuclear modification factors 
{\it R}$_{{\it d}{\text {Au}}}$,
but results in a somewhat better agreement with data (solid histograms).
Similar results are obtained for minimum bias data.
The data indicate at most a small variation  
with centrality of the factor {\it f}  consistent with 
the broadening originating at the parton level.

\begin{figure} [h]
\centering
\includegraphics[width=0.8\linewidth]{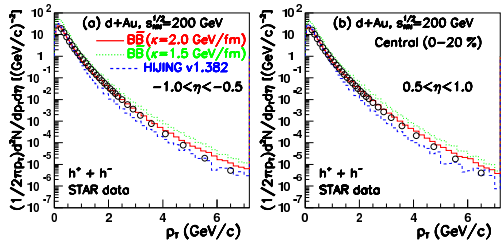}
\caption[p_t,rdAu,h+ + h- ] {\small (Color online)
{\small HIJING/B\=B v2.0} (with SCF) and {\small HIJING v1.382} 
predictions for $p_T$ spectra of total inclusive charged hadrons, 
for central (0--20\%) {\it d}+Au collisions,
in backward ( $-1.0 < \eta < -0.5$, left panel)
and forward ($0.5 < \eta < 1.0$, right panel) pseudo-rapidity region.
The calculations assume $\kappa_2$=2.0 GeV/fm for {\it d}+Au 
(solid histograms; Tab. II, Set 4) and $\kappa_1$=1.5 GeV/fm for 
dotted histograms; Tab. II, Set 3). The dashed histograms are the results 
obtained within {\small HIJING v1.382}, with default parameters.
The calculations in both models are obtained with shadowing and without
jet quenching. The data are from STAR \cite{star04_asym}.
Only statistical error bars are shown.
\label{fig:fig6}
}
\end{figure}

 A slight discrepancy is seen in the description of
observed $p_T$ spectra at midrapidity for total inclusive charged hadrons
yields in central (0--20 \%) {\it d}+Au collisions.
In order to understand this,
we analyse charged hadron asymmetries (i.e., the ratio of the hadron yield 
in backward rapidity to forward rapidity intervals) as 
proposed in reference \cite{star04_asym}.
Since this observable is defined as a ratio, it is important 
to check the results for both numerator and denominator.
Fig.~\ref{fig:fig6} presents the yields in different 
pseudo-rapidity interval, backward (part a) and forward (part b),
obtained within {\small HIJING/B\=B v2.0} and {\small HIJING} models.
The absolute $p_T$ distribution 
are better described by {\small HIJING/B\=B}  
in comparison to the results of regular {\small HIJING}.

\begin{figure} [hbt!]
\centering
\includegraphics[width=0.8\linewidth]{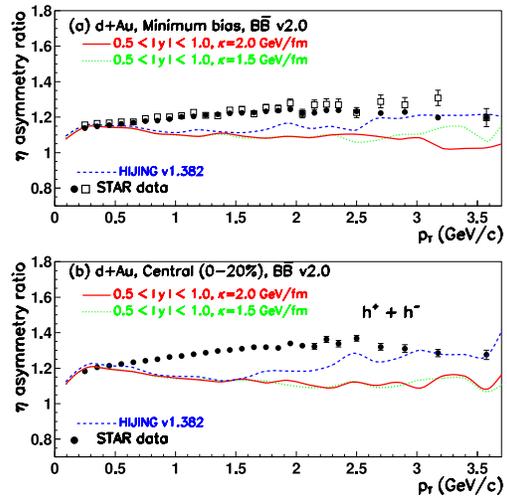}
\caption[p_t,rdAu,h+ + h- ] {\small (Color online)
The ratio of charged hadron spectra in the backward 
rapidity to forward rapidity region for minimum bias 
(upper panel) and central (0--20\%) (lower panel)
{\it d}+Au collisions. The curves have the same meaning as 
in Fig.~\ref{fig:fig6}. The data are from STAR \cite{star04_asym}.
Only statistical error bars are shown.
\label{fig:fig7}
}
\end{figure}

In Figure~\ref{fig:fig7}, we analyse the observed charged 
hadron asymmetries within both models 
for central (0--20\%) and minimum bias {\it d}+Au collisions.
{\small HIJING/B\=B} and {\small HIJING} models predict similar asymmetries. 
The calculated asymmetries are greater 
than 1.0, in qualitative agreement with the observations and contrary 
to pQCD predictions \cite{xnwang_asym03}.
Within our model, the results on asymmetry are also little sensitive to the 
strength of the assumed longitudinal chromoelectric field.
For central (0--20\%) {\it d}+Au collisions, 
increasing string tension results in a decrease of the absolute yield
(see Fig.~\ref{fig:fig6}), without changing the asymmetry significantly.
There are still some differences between the $p_T$ dependence of
the predicted and measured asymmetry.
A relatively good agreement is obtained for minimum bias 
{\it d}+Au collisions in the region ($0.0 < |\eta| < 0.5$), not shown here.
This observable is very delicate and great care should be
taken on its interpretation.

Our main conclusion is that the small asymmetry is not well reproduced by
either models. 
These observed differences may be related also   
to the disagreement in the calculated $p_T$ spectra of identified particles 
for central {\it d}+Au collisions discussed above (Sec. IIIA). 
  
\begin{figure} [h]
\centering
\includegraphics[width=0.8\linewidth]{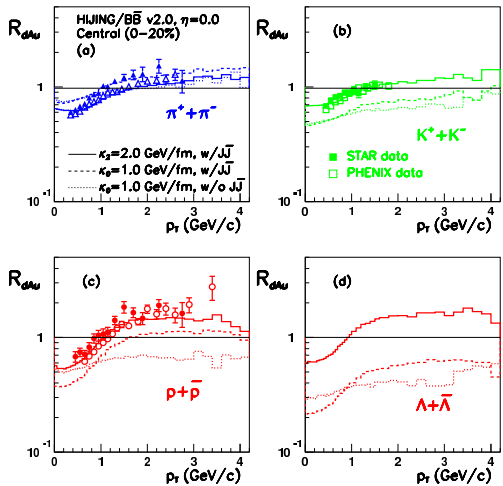}
\vskip 0.5cm

\caption[Id r_dAu, 0-20] {\small (Color online)
{\small HIJING/B\=B v2.0} predictions for species dependence of 
nuclear modification factor ({\it R}$_{{\it d}{\text {Au}}}$)
in central (0--20\%) {\it d}+Au collisions at $\sqrt{s_{NN}}$=200 GeV.
Shown are results for:
(a) charged pions, (b) kaons, (c) inclusive {\it p}+$\bar{\it p}$, and
(d) inclusive $\Lambda+\bar{\Lambda}$. 
The histograms have the same meaning as in 
right panel of Fig.~\ref{fig:fig5}.
The data are from STAR (filled symbols) \cite{Adams:2006_1} and 
from PHENIX (open symbols) \cite{Adler:2006xd}.   
Only statistical error bars are shown.
\label{fig:fig8}
}
\end{figure}

Higher sensitivity to the new dynamics implemented 
in the {\small HIJING/B\=B v2.0}
is obtained by study of nuclear modification factors of identified particles.
To better quantify possible effects of strong color 
field on particle production, results of {\it R}$_{{\it d}{\text {Au}}}(p_T)$ 
for central ($<N_{\text {bin}}>$=14.8) {\it d}+Au collisons  
where higher sensitivity to SCF is expected, 
are presented in  Fig.~\ref{fig:fig8}.
Because of their dominance, the production of pions 
is only moderately modified when we assume  
an increase of the string tension value since the total energy is conserved.
Taking into account SCF effects (solid histograms)
results in changes at moderate $p_T$ 
of $\approx$ 15--20\% of the predicted pion yield
(Fig.~\ref{fig:fig8}a).  
The pions yield in central {\it d}+Au collisions is enhanced relative 
to {\it p}+{\it p} collisions 
(i.e.,  {\it R}$_{{\it d}{\text {Au}}}^{\pi}(p_T)>$ 1)
an effect reproduced by our calculation.
The scaling behavior of sum of protons and anti-protons   
({\it p}+$\bar{\it p}$) is somewhat different than
that of the pions. The (anti)protons are much more 
sensitive to $J\bar{J}$ loops and SCF effects, and as a result 
{\it R}$_{{\it d}{\text {Au}}}^{(p+\bar{p})}(p_T)$ is predicted to be higher
than that of pions at moderate $p_T$. 
This result is consistent with the data of Ref.~\cite{Adams:2003qm},
where this behavior was explained by different Cronin enhancement 
for pions and protons. 
However, within our model it is not only the ``Cronin effect'' which 
modified {\it R}$_{{\it d}{\text {Au}}}$, 
but an interplay of initial state effects
with an important contribution coming from $J\bar{J}$ loops and SCF.
In contrast to pions, (multi)strange particles show a high sensitivity to 
the presence of SCF. Kaons (Fig.~\ref{fig:fig8}b) and
lambda particles (Fig.~\ref{fig:fig8}d), show an increase at moderate $p_T$
by a factor of approximately 1.5 and 3.0 respectively, 
relative to the calculation that do not include  
$J\bar{J}$ loops and SCF effects (dotted histograms).
As opposed to pions such an increase results in a predicted strong enhancement 
of the lambda yield relative to scaled binary collisions. 

\begin{figure} [hbt!]
\centering
\includegraphics[width=0.8\linewidth]{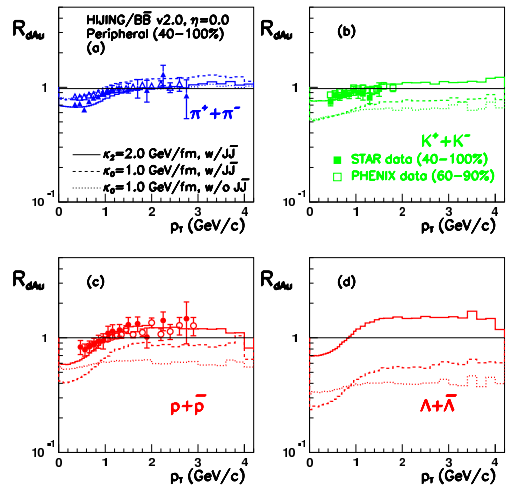}
\caption[ID r_dau 40-100] {\small (Color online)
{\it R}$_{{\it d}{\text {Au}}}$ for peripheral 
(40--100\%) {\it d}+Au collisions 
at $\sqrt{s_{NN}}$=200 GeV. The histograms have the same 
meaning as in right panel of Fig.~\ref{fig:fig5}.
The data are from STAR (filled symbols) \cite{Adams:2006_1}
and from PHENIX (open symbols) \cite{Adler:2006xd}.
Only statistical error bars are shown.
\label{fig:fig9}
}
\end{figure}

Figure~\ref{fig:fig9} presents the values of 
{\it R}$_{{\it d}{\text {Au}}}$ for 
peripheral (40--100\%) collisions, where the mean number
of collisions is very small,  $<N_{\text {bin}}>$=4.5.
Our calculations shows that SCF effects play an 
important role already 
when there is only a few number of binary collisions.
The calculation including SCF describe well the experimental 
results.

\begin{figure} [h]
\centering
\includegraphics[width=0.8\linewidth]{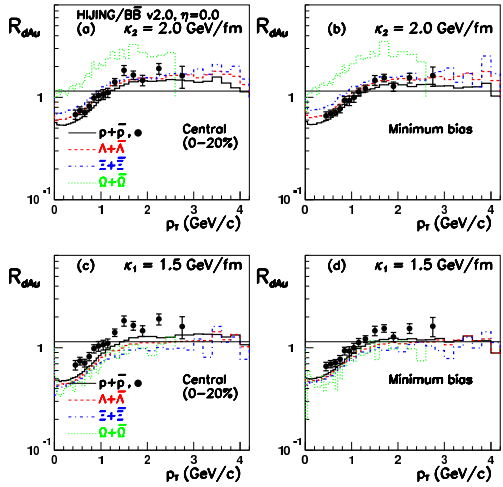}
\caption[R_dAu sensitivity, k] {\small (Color online)
{\small HIJING/B\=B v2.0} predictions including SCF 
with $\kappa_2$=2.0 GeV/fm (Tab. II, Set 4) (upper panels)
and with $\kappa_1$=1.5 GeV/fm (Tab. II, Set 3) (lower panels)
for central (0--20\%) (left panels) and minimum bias (right panels) 
{\it d}+Au collisions at $\sqrt{s_{NN}}$=200 GeV.
Shown are results for: inclusive {\it p}+$\bar{\it p}$ (solid),
inclusive $\Lambda+\bar{\Lambda}$ (dashed),
$\Xi^++\Xi^-$ (dash-dotted) and $\Omega^++\Omega^-$ (dotted histograms).
The data are from STAR \cite{Adams:2006_1}.
Only statistical error bars are shown.
\label{fig:fig10}
}
\end{figure}

The sensitivity to transient field fluctuations
is shown in Fig.~\ref{fig:fig10} which present 
the predicted {\it R}$_{{\it d}{\text {Au}}}^{\pi}(p_T)$ assuming 
$\kappa_2$=2.0 GeV/fm and $\kappa_1$=1.5 GeV/fm for 
 central (0--20\%) {\it d}+Au collisions.
For $\kappa_1$=1.5 GeV/fm the calculation predicts
a value of {\it R}$_{{\it d}{\text {Au}}}$ close to one and that decrease 
somewhat with the mass and strangeness content.
Increasing $\kappa$ to $\kappa_{2}$=2.0 GeV/fm
results in a larger enhancement relative to {\it p}+{\it p} 
(i.e., {\it R}$_{{\it d}{\text {Au}}} >$ 1)
and a reversing of the order of the nuclear modification factor
as a function of the mass and strangeness content.
In particular, the model predicts an enhancement in the multi-strange 
(anti)hyperon production at moderate $p_T$, by a factor 
of roughly 2 relative to binary scaling 
for $\Xi$s and by a factor of approximately 3 for $\Omega$s.

Preliminary data on hyperons $\Lambda$ and $\Xi$
\cite{rene_04,mironov_sqm04}, 
seems to favour $\kappa_{1}$=1.5 GeV/fm. 
However, due to large statistics and systematics error, 
no clear conclusion could be drawn.
Measurements of $\Omega$ particles could help us
to draw a final conclusion on the importance of  
SCF fluctuations at top RHIC energy and  to determine the effective
value ($\kappa$) of string tension, in {\it d}+Au collisions.

\begin{figure} [hbt!]
\centering
\includegraphics[width=0.8\linewidth]{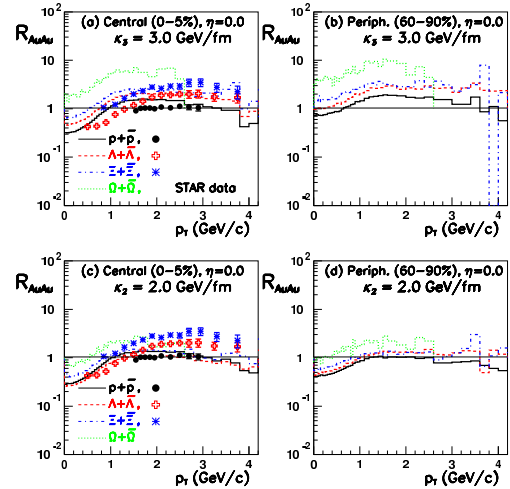}
\caption[R_AuAu, sensitivity, k ] {\small (Color online)
{\small HIJING/B\=B v2.0} predictions including SCF 
with $\kappa_3$=3.0 GeV/fm (Tab. II, Set 5) (upper panels),
and with $\kappa_2$=2.0 GeV/fm (Tab. II, Set 4) (lower panels),
for central (0--5\%) (left panels) and peripheral (60--90\%) (right panels) 
Au+Au collisions at $\sqrt{s_{NN}}$=200 GeV.
The histograms have the same meaning as in Fig.~\ref{fig:fig10}.
The data are from STAR \cite{Adams:2006_1,Adams:2006_2,Adams:2006ke}.
Only statistical error bars are shown.
\label{fig:fig11}
}
\end{figure}

We show in Sec. IIIA, that a new parametrization of the 
{\it p}+{\it p} interaction is necessary to describe the identified 
particle spectra especially for strange and multi-strange
particles. This parameterization is different than that of {\small PYTHIA} 
mainly in assuming different constituent and current
quark masses for diquark. Therefore the strength of SCF (and $\kappa$)
introduced in our previous work \cite{prc72_top05} to best described 
{\it R}$_{{\text {AuAu}}}$ has to be modified because 
of this new {\it p}+{\it p} baseline.
Figure~\ref{fig:fig11} show the sensitivity of NMF {\it R}$_{{\text {AuAu}}}$
to transient color field fluctuations for the {\it p}+$\bar{\it p}$ yield  
and for (multi)strange particles.
The results assuming $\kappa$=$\kappa_2$=2.0 GeV/fm and 
$\kappa$=$\kappa_3$=3.0 GeV/fm are shown for both central (0--5\%) 
and peripheral (60--90\%) Au+Au collisions.
In agreement with the data,  
the calculated {\it p}+$\bar{\it p}$ yield scale approximately
with the number of binary collisions.
In contrast, the observed {\it R}$_{{\text {AuAu}}}$ 
for (multi)strange particles 
show an enhancement over {\it p}+$\bar{\it p}$, which increase with 
the mass and strangeness content.
A good descriptions of NMF {\it R}$_{{\text {AuAu}}}$
for (multi)strange at moderate $p_T$, is obtained only for the largest value 
of $\kappa$, i.e., $\kappa_3$ (Fig.~\ref{fig:fig11}a).
A clear ordering with strangeness content, seen in the data, is not reproduced
by lower value $\kappa$=$\kappa_2$=2.0 GeV/fm (Fig.~\ref{fig:fig11}c). 
This clearly illustrates that in Au+Au collisions, 
fluctuations of the chromoelectric field are higher than in 
{\it d}+Au collisions, and affect especially (multi)strange particle yields.
The {\it p}, $\pi^+$, {\it K}$^+$, and their antiparticles are less affected
and their yield could also be well described with 
$\kappa$=$\kappa_2$=2.0 GeV/fm (see also Fig.~\ref{fig:fig4}a,b).
One note that very similar enhancements are predicted for 
central and peripheral collisions, with 
slightly higher value predicted for peripheral (60--90\%) 
collisions, that could be attributed to different strength of 
quenching effect.

In contrast to {\it R}$_{{\text {AuAu}}}$, {\it R}$_{{\text {CP}}}$ obtained 
as scaled ratios of central (0--5\%) and peripheral 
(60--90 \%) Au+Au collisions show for (multi)strange particles   
a slight suppression relative to binary collisions scaling
 ({\it R}$_{{\text {CP}}}$), see Fig. 3b,d in Ref.~\cite{prc72_top05}.
This suppression is mainly due to a final state effect,
``jet quenching'', which is stronger in central than in   
peripheral collisions. 
A striking difference between {\it R}$_{{\text {AuAu}}}$ and  
{\it R}$_{{\text {CP}}}$ has been reported by STAR \cite{rene_04}. 
This experimental fact could be
explained in our model as a consequence of interplay of 
initial and final state effects.  
Baryon junction loops ($J\bar{J}$) and SCF effects 
are taken into consideration in both central and peripheral collisions.
In {\it p}+{\it p} collisions the contribution of junction loops 
($J\bar{J}$) is negligible (due to small probability and kinematical cuts), 
and SCF have  a reduced strength ($\kappa_1$=1.5 GeV/fm).
This show that there could be a significant difference in the 
value and the meaning of {\it R}$_{{\text {CP}}}$ and 
{\it R}$_{{\text {AuAu}}}$ due to differences in the baseline 
used for comparison with binary scaling, i.e., peripheral (Au+Au) 
yields for {\it R}$_{{\text {CP}}}$ and {\it p}+{\it p} yields 
for {\it R}$_{{\text {AuAu}}}$.

\section{Summary and Conclusions}

We studied the influence of possible 
multi-gluon dynamics ({\em gluon junctions}) and 
strong longitudinal color fields (SCF)
on particle production in heavy-ion collisions.
A new parameterization of the {\it p}+{\it p} interaction based on 
new constituent and current quark masses for diquark
is introduced. It leads to a simultaneously good description of non-strange 
and strange sector in {\it p}+{\it p} collisions.
We show that $J\bar{J}$ loops play an important role in 
particle production at mid-rapidity in {\it d}+Au and Au+Au colisions at RHIC
energies. Introducing a new $J\bar{J}$ loops algorithm
in the framework of  {\small HIJING/B\=B v2.0}, 
leads to a consistent and significant improvement in the description 
of the recent experimental results for protons, pions, and kaons 
for both reactions.
The present studies within our model are limited to
the effect of initial state baryon production via possible junction dynamics
in strong fields. It would be very interesting 
to consider a generalization of back reaction effects \cite{kluger_back92}
to the case not only of pair production relevant for mesons 
but to the more difficult
three string junction configurations needed to describe baryon production.
Baryon productions via the conventional default quark-diquark mechanisms in
the Lund string model are known to be inadequate even in 
{\it e}$^+$+{\it e}$^-$ phenomenology. 
This is one of the main reasons for
our continued exploration of baryon junction alternative mechanisms.

The strange and multi-strange particles could only be
described in the framework of string models if we consider
SCF effects. The mechanisms of their production 
is very sensitive to the early phase of nuclear 
collisions, when fluctuations in the color field strength are highest.
Within {\small HIJING/B\=B v2.0}, SCF effects 
are modeled by varying the effective string tension that controls the
{\it q}$\bar{\it q}$ and qq$\overline{\text {qq}}$ pair creation rates
and strangeness suppression factors.
The mid-rapidity yield of (multi)strange particles  
favor a large value of the   
average string tension for both collisions 
(i.e., $\kappa\, \approx\,$ 2.0 GeV/fm in {\it d}+Au, and 
$\kappa \,\approx\, $ 3.0 GeV/fm in Au+Au).
A strong enhancement in strange baryon nuclear modification factors 
{\it R}$_{{\text {AA}}}$ with increasing strangeness content is predicted for  
{\it d}+Au and Au+Au.  
In contrast, a clear ordering with strangeness content is not predicted 
for lower mean values of string tension. 

A greater sensitivity to SCF effects is predicted 
for the nuclear modification factors of multi-strange 
hyperons $\Xi$ and $\Omega$.
In particular, the measurement of $\Omega$ and $\bar\Omega$ yields would
provide an important test of the relevance of SCF 
fluctuations, helping us to choose appropiate values for the suppression 
factors $\gamma_Q$ (where Q=qq, us, {\it s}, {\it u}), which   
have strong dependence on the main parameters of QCD 
(the constituent and current quark masses) and on the system size. 
Even though the success of this procedure has been clearly illustrated here,
the full understanding
of the production of (multi)strange particles in
relativistic heavy-ion collisions remain an exciting open question,
and challenge many theoretical ideas.

\section{Acknowledgments}

We thank N. Xu and Helen Caines for helpful discussions
and suggestions throughout this project. It is a pleasure to thank Julia
Velkovska for useful discussions.

This work was partly supported by the Natural Sciences and 
Engineering Research Council of Canada and the 
Fonds Nature et Technologies of Quebec.  
This work was supported also by the Director,
Office of Energy Research, Office of High Energy and Nuclear Physics,
Division of Nuclear Physics, and by the Office of Basic Energy
Science, Division of Nuclear Science, of the U. S. Department 
of Energy under Contract No. DE-AC03-76SF00098 and
DE-FG02-93ER-40764. One of us (MG), gratefully acknowledges partial 
support also from FIAS and GSI.

\newpage

\appendix*
\section{Tables}
\begin{table}[h]
\caption{Main parameters used in the calculation 
for {\it p}+{\it p} collisions.
The parameters are defined in the text. Set 1 is used in {\small PYTHIA}
(and regular {\small HIJING}). Set 2 is obtained using the constituent quark
masses from Sec. II B. Set 3 includes an additional increase of the string
tension to  $\kappa_1$ = 1.5 GeV/fm.}
\begin{tabular}{ccccccccc} \hline \hline  
{\it p} + {\it p} &  $\kappa$ (GeV/fm) & $\gamma_{{\text {qq}}}$ & 
 $\gamma_{s}$ &  $\gamma_{{\text {us}}}$ & $\gamma_{10}$ & 
$\sigma_{q}$ (GeV/{\it c}) & {\it f} & \\
\hline
 Set 1 & $\kappa_0$ = 1.0 & 0.09 & 0.30 & 0.40 & 0.05 & 0.350 & 1 & \\
Set 2 & $\kappa_0$ = 1.0 & 0.02& 0.30 & 0.40 & 0.05 & 0.350 & 1 & \\
Set 3 & $\kappa_1$ = 1.5 & 0.07 & 0.45 & 0.54 & 0.09 & 0.430 & 1 & \\
\hline
\hline
 \end{tabular}
\label{tab:tab1}
 \end{table}

\begin{table}[h]
\caption{Main parameters used in {\it d}+Au and Au+Au collisions. 
The parameters are defined in the text. 
Set 1 correspond to calculations without $J\bar{J}$ loops
and SCF effects. Set 2 adds the contribution of $J\bar{J}$ loops.
Set 3--5 include both effects and correspond to different values
of the string tension.}
\begin{tabular}{ccccccccc}    \hline \hline
({\it d})Au + Au &  $\kappa$ (GeV/fm) &  $\gamma_{{\text {qq}}}$ & 
$\gamma_{s}$ & $\gamma_{{\text {us}}}$ & $\gamma_{10}$ & 
{\bf $\sigma_{\it q}$} (GeV/{\it c}) & {\it f} & \\
\hline
Set 1 & $\kappa_0$ = 1.0 & 0.02 & 0.30 & 0.40 & 0.05 & 0.350 & 1 & \\
Set 2 & $\kappa_0$ = 1.0 & 0.02 & 0.30 & 0.40 & 0.05 
& 0.350 & 3 & \\
Set 3 & $\kappa_1$ = 1.5 & 0.07 & 0.45 & 0.54 & 0.09 & 0.430 & 3 & \\
Set 4 & $\kappa_2$ = 2.0 & 0.14 & 0.55 & 0.63 & 0.12 & 0.495 & 3 & \\
Set 5 & $\kappa_3$ = 3.0 & 0.27 & 0.67 & 0.74 & 0.18 & 0.606 & 3 & \\
\hline
\hline
 \end{tabular}
\label{tab:tab2}
 \end{table}


\begin{thebibliography}{199}







\bibitem{Adcox:2001mf} PHENIX Collaboration,~K.~Adcox 
{\it et al.},~Phys.~Rev.\\
~Lett.~{\bf 89}, 092302 (2002).


\bibitem{Adler:2003kg} PHENIX Collaboration,~S.~S.~Adler {\it et al.},
~Phys.~Rev.\\
~Lett. {\bf 91}, 172301 (2003). 

\bibitem{Vitev:2001zn}
I.~Vitev and M.~Gyulassy,
Phys.\ Rev.\ C {\bf 65}, 041902 (2002).

\bibitem{Gyulassy:2003mc}~M.~Gyulassy,~I.~Vitev,~X.~-N.~Wang
and~B.~W.~Zhang, Quark Gluon Plasma 3, ed. by~R.~C.~Hwa and~X.~-N.~Wang 
(World Scientific, Singapore, 2003), pp.123-191.


\bibitem{Adcox:2001jp}PHENIX Collaboration,~K.~Adcox {\it et al.},~Phys.~Rev.\\
~Lett.{\bf 88}, 022301 (2002).

\bibitem{David:2001gk} PHENIX Collaboration,~G.~David {\it et al.},
~Nucl.~Phys. {\bf A698}, 227 (2002).






\bibitem{prc70_top04}~V.~Topor~Pop,~M.~Gyulassy,~J.~Barrette,~C.~Gale,
X.~-N.~Wang, and~N.~Xu, 
~Phys.~Rev.~C {\bf 70}, 064906 (2004), and references therein.

\bibitem{prc72_top05}~V.~Topor~Pop,~M.~Gyulassy,~J.~Barrette,~and~C.~Gale,
~Phys.~Rev.~C {\bf 72}, 054901 (2005). 



\bibitem{Kharzeev:1996sq}
  D.~Kharzeev,
  Phys.\ Lett.\ B {\bf 378}, 238 (1996).



\bibitem{muller03}~R.~J.~Fries,~B.~Muller,~C.~Nonaka, and
~S.~A.~Bass, Phys. Rev. C {\bf 68}, 044902 (2003);
~V. Greco,~C.~M. Ko, and~P.~Levai,~Phys.~Rev.~Lett. {\bf 90}, 202302 (2003);
~R.~C.~Hwa and ~C.~B.~Yang,~Phys.~Rev. C {\bf 70}, 024905 (2004).

\bibitem{miklos_zf_91} ~M.~Grabiak,~J.~A.~Casado, and~M.~Gyulassy,
~Z.~Phys. C {\bf 49}, 283 (1991).

\bibitem{biro_91}~T.~S.~Biro,~B.~M\"uller, and~X.~-N.~Wang,
~Phys.~Lett.~B {\bf 283}, 171 (1992).

\bibitem{heinz_03}~P. F. Kolb and U. Heinz,
ibid. \cite{Gyulassy:2003mc}, pp634-714 (2003).

\bibitem{Biro:1984cf}
  T.~S.~Biro, H.~B.~Nielsen and J.~Knoll,
  Nucl.\ Phys.\ {\bf B245}, 449 (1984).

\bibitem{soff_jpg04}~S.~Soff,~J.~Phys. G {\bf 30},
 S139 (2004);~S.~Soff,~J.~Randrup,~H.~St\"ocker, and~N.~Xu, 
~Phys.~Lett.~B {\bf 551}, 115 (2003);~S.~Soff,~S.~Kesavan,
~J.~Randrup,~H.~St\"ocker, and~N.~Xu, J. Phys. G {\bf 30}, L35 (2004).


\bibitem{kapusta05_1}~R.~J.~Fries,~J.~I.~Kapusta and~Y.~Li,
hep-ph/0511101; nucl-th/0604054.

\bibitem{lappi06_1}~T.~Lappi~and~L.~McLerran,~Nucl.~Phys. 
{\bf A772}, 200 (2006).

\bibitem{asakawa06_1}~M.~Asakawa,~S.~A.~Bass and~B.~Muller,
hep-ph/0603092.


\bibitem{ripka_03}~G.~Ripka, Lecture Notes in Physics, {\bf 639}
(2004), (Ed. Springer Verlag, Berlin, Germany).


\bibitem{rafelski_82}~J.~Rafelski~and~B.~M\"uller,
~Phys.~Rev.~Lett. {\bf 48}, 1066 (1982);~{\bf 56},~2334E~(1986);\\
~H.~Z.~Huang~and~J.~Rafelski,~AIP~Conf.~Proc.~{\bf 756}, 210 (2005).

\bibitem{rene_04}~R.~Bellwied,~J.~Phys.~G {\bf 30}, S29 (2004).
\bibitem{caines_jpg05}~H.~Caines,~J.~Phys.~G {\bf 31}, S1057 (2005).

\bibitem{cgreiner_02}~C.~Greiner,~J.~Phys.~G {\bf 28}, 1631 (2002).

\bibitem{armesto96}~N.~Armesto,~M.~A.~Braun, E.~C.~Ferreiro, and
C.~Pajares, Phys. Lett. B {\bf 344}, 301 (1995);
~N.~Armesto,~M.~A.~Braun, E.~C.~Ferreiro,~C.~Pajares, and~Yu.~M.
Shabelski, Phys. Lett. B {\bf 389}, 78 (1996). 

\bibitem{antai97}~Tai~An and~Sa~Ben-Hao,~Phys.~Lett.~B {\bf 409}, 
393 (1997). 


\bibitem{csernai01}~V.~Magas,~L.Csernai, and~D.~Strottman,
~Phys.~Rev.~C {\bf 64}, 014901 (2001).

\bibitem{Vance:1999pr}
  S.~E.~Vance and M.~Gyulassy,
  Phys.\ Rev.\ Lett.\  {\bf 83}, 1735 (1999).

\bibitem{svance99}~S.~E.~Vance,~J.~Phys.~G {\bf 27}, 603 (2001).

\bibitem{hij_top06}~The updated Fortran sources of the program can be\\
 downloaded from URL address:\\  
{\em http://www.physics.mcgill.ca/$\sim$toporpop}. 

\bibitem{wang_97}~M.~Gyulassy,~X.~-N.~Wang,~Phys.~Rev.~D {\bf 44}, 3501 (1991);
~M.~Gyulassy,~X.~-N.~Wang,~Comput.~Phys.~Commun., {\bf 83},
307 (1994);~X.~-N.~Wang, Phys. Rep. {\bf 280}, 287 (1997).


\bibitem{pyt_94}~T.~Sj\"ostrand, Comput. Phys. Commun., 74 (1994).

\bibitem{prc68_top03}~V.~Topor~Pop,~M.~Gyulassy,~J.~Barrette, 
C.~Gale,~X.~-N.~Wang,~N.~Xu and~K.~Filimonov, 
~Phys.~Rev.~C {\bf 68}, 054902 (2003).


\bibitem{svance98}~S.~E.~Vance,~M.~Gyulassy, and~X.~-N.~Wang, 
Phys.~Lett.~B {\bf 443}, 45 (1998).


\bibitem{castillo_04} STAR Collaboration,~J.~Adams {\it et al.,}
~Phys.~Rev.~Lett. {\bf 95}, 122301 (2005).

\bibitem{collins_77}~P.~D.~Collins, {\em An Introduction to
Regge Theory and High Energy Physics}, Cambridge Univ. Press, 1977.


\bibitem{kopelovic99}~B.~Z.~Kopeliovich and~B.~Povh,
~Phys.~Lett.~Phys.~Lett. B {\bf 446}, 321 (1999).


\bibitem{zabrodin_04}~A.~V.~Prozorkevich,~S.~A.~Smolyansky,~V.~V.~Skokov, and
~E.~E.~Zabrodin,~Phys.~Lett.~B {\bf 583}, 103 (2004).

\bibitem{skokov_prd05}~V.~V.~Skokov, and~P.~Levai,~Phys.~Rev. 
D {\bf 71}, 094010 (2005).






\bibitem{soff_99}~S.~Soff,~S.~A.~Bass,~M.~Bleicher,~L.~Bravina,
E.~Zabrodin,~H. St\"ocker, and~W.~Greiner,
Phys.~Lett. B {\bf 471}, 89 (1999).

\bibitem{brown_91}~G.~E.~Brown and~M.~Rho, 
~Phys.~Rev.~Lett. {\bf 66}, 2720 (1991). 

\bibitem{bleicher_01}~M.~Bleicher {\it et al.},
Phys. Rev. C {\bf 64}, 011902 (2001).   

\bibitem{bielich_04}~J.~Schaffner-Bielich, J. Phys. G {\bf 30}, R245
(2004).

\bibitem{rische_05}~S.~B.~R\"uster,~V.~Werth,~M.~Buballa,
~I.~A.~Shovkovy,\\
~and~D.~H.~Rischke,~Phys.~Rev.~D~{\bf 72},~034004~(2005).

\bibitem{dima_k05}~D.~Kharzeev, and~K.~Tuchin, 
~Nucl.~Phys. {\bf A753}, 316 (2005). 

\bibitem{nussinov_80}~A.~Casher,~H.~Neuberger,~A.~Nussinov,
~Phys.~Rev. D {\bf 20}, 179 (1979);~Phys.~Rev. D {\bf 21}, 1966 (1980). 


\bibitem{schwinger_51}~J.~S.~Schwinger,~Phys.~Rev.~{\bf 82},
664 (1951). 

\bibitem{gc_nayak06}
~G.~C.~Nayak, P. van Nieuwenhuizen,~Phys.~Rev.~D {\bf 71}, 125001,2005; 
~G.~C.~Nayak,~Phys.~Rev. D {\bf 72}, 125010 (2005);
~F.~Cooper~and~G.~C.~Nayak,~Phys.~Rev.~D {\bf 73}, 065005 (2006).

\bibitem{gies_prd05}~H.~Gies and K.~Klingm\"uller,
~Phys.~Rev. D {\bf 72}, 065001 (2005).

\bibitem{fried_prd06}~H.~M.~Fried and ~Y.~Gabellini,
~Phys.~Rev. D {\bf 73}, 011901(R) (2006).

\bibitem{kluger_back92}~Y.~Kluger,~J.~M.~Eisenberg,~B.~Svetitsky,
~F.~Cooper,~and~E.~Mottola,~Phys.~Rev.~D {\bf 45}, 4659 (1992);
~Y.~Kluger,~J.~M.~Eisenberg,~B.~Svetitsky,~F.~Cooper,~and~E.~Mottola,
~Phys.~Rev.~Lett. {\bf 67}, 2427 (1991);
~Y.~Kluger,~E.~Mottola, and~J.~Eisenberg,
~Phys.~Rev.~D {\bf 58}, 125015 (1998).

\bibitem{cooper_plb03}~F.~Cooper,~E.~Mottola,~and~G.~C.~Nayak,
~Phys.~Lett.`B {\bf 555}, 181 (2003).

\bibitem{pdb_04}~E.~Eidelman {\it et al.}, (Particle Data Group),
~Phys.~Lett.~B {\bf 592}, 1 (2004). 

\bibitem{ripka_04_1}M.~Cristoforetti,~P.~Faccioli,~G.~Ripka, 
and~M.~Traini,\\
~Phys.~Rev.~D {\bf 71}, 114010 (2005).

\bibitem{amelin_01}~N.~S.~Amelin,~N.~Armesto,~C.~Pajares, and~D.~Sousa, 
Eur. Phys. J. C {\bf 22}, 149 (2001).

\bibitem{flork_04}~W.~Florkowski, Acta~Phys.~Polon. {\bf B35},
799 (2004).
\bibitem{raf_06}~S.~Steinke and J. Rafelski, nucl-th/0607066.


\bibitem{hoft_04}~G.~Hooft, hep-th/0408148, 
Presented at the Workshop on Hadrons and Strings, Trento, July, (2004).

\bibitem{cgreiner_04_y}~G.~Martens,~C.~Greiner,~S.~Leupold,
and~U.~Mosel,\\
~Phys.~Rev.~D {\bf 70}, 116010 (2004).

\bibitem{takahashi_05}~T.~T.~Takahashi,~H.~Matsufuru,~Y.~Nemoto, 
and~H.~Suganuma,~Phys.~Rev.~Lett. {\bf 86}, 18 (2001);
~T.~T.~Takahashi,~H.~Suganuma,~Y.~Nemoto, and~H.~Matsufuru,
~Phys.~Rev. D {\bf 65}, 054503 (2002);
~F.~Okiharu,~H.~Suganuma and~T.~T.~Takahashi,
~Phys.~Rev. D {\bf 72}, 014505 (2005).










\bibitem{enteria_05jpg}~D.~d'Enteria, ~J.~Phys.~G~{\bf 31},~S491~(2005).



\bibitem{Adams:2003kv} STAR Collaboration,~J.~Adams {\it et al.},
~Phys.~Rev.~Lett. {\bf 91}, 172302 (2003).



\bibitem{Adler:2003qi} PHENIX Collaboration,~S.~S.~Adler {\it et al.},~Phys.\\
~Rev.~Lett. {\bf 91}, 241803 (2003).




\bibitem{Adams:2003xp}~STAR~Collaboration,~J.~Adams {\it et al.},
  Phys.\ Rev.\ Lett.\  {\bf 92}, 112301 (2004).

\bibitem{Adams:2003am} STAR Collaboration,~J.~Adams {\it et al.},
~Phys.~Rev.~Lett. {\bf 92}, 052302 (2004). 




\bibitem{Adams:2003qm}
~STAR~Collaboration,~J.~Adams {\it et al.},~Phys.~Lett. B {\bf 616}, 8 (2005).

\bibitem{Adams:2006_1}~STAR~Collaboration,~J.~Adams {\it et al.},~Phys.~Lett. B {\bf 637}, 161 (2006). 

\bibitem{Adams:2006_2} ~STAR~Collaboration,~J.~Adams {\it et al.},
nucl-ex/0607033 (submitted to Phys. Rev. C)





\bibitem{Bellwied:2005bi} STAR Collaboration, ~R.~Bellwied {\it et al.},
nucl-ex/0511006.








\bibitem{Adler:2003au} PHENIX Collaboration,~S.~S.~Adler {\it et al.},
~Phys.~Rev.~C {\bf 69}, 034910 (2004).



\bibitem{Adams:2005dq} STAR Collaboration,
~J.~Adams {\it et al.},~Nucl.~Phys.~{\bf A757}, 102 (2005).

\bibitem{Adams:2006ke} STAR Collaboration,~J.~Adams {\it et al.},
~nucl-ex/0606014 (submitted to~Phys.~Rev.~Lett.).





\bibitem{Adler:2006xd}
  PHENIX Collaboration, S.~S.~Adler {\it et al.},
  nucl-ex/0603010 (submitted to Phys. Rev. C).



\bibitem{alexopoulos98} ~T.~Alexopoulos~{\it et al.}, 
~Phys.~Lett.~B~{\bf 435}, 453 (1998).

\bibitem{Adler:2006sc}~PHENIX~Collaboration,
  S.~S.~Adler {\it et al.}, hep-ex/0605039.

\bibitem{feynman_77}~R.~P.~Feynman,~R.~D.~Field, and~G.~C.~Fox,
~Nucl.~Phys.~{\bf B128}, 1 (1977).

\bibitem{kniehl_01npb} ~B.~Kniehl,~G.~Kramer and~B.~Potter,
~Nucl.~Phys. {\bf B597}, 337 (2001).




\bibitem{Adler:2003ii}~PHENIX~Collaboration, S.~S.~Adler {\it et al.},
  Phys.\ Rev.\ Lett.\  {\bf 91}, 072303 (2003).


\bibitem{Adams:2003im}~STAR~Collaboration, J.~Adams {\it et al.},
  Phys.\ Rev.\ Lett.\  {\bf 91}, 072304 (2003).


\bibitem{Arsene:2003yk}~BRAHMS~Collaboration, I.~Arsene {\it et al.},
  Phys.\ Rev.\ Lett.\  {\bf 91}, 072305 (2003).

\bibitem{Back:2003ns}~PHOBOS~Collaboration, B.~B.~Back {\it et al.},
  Phys.\ Rev.\ Lett.\  {\bf 91}, 072302 (2003).

\bibitem{cronin_75}~J.~W.~Cronin {\it et al.},~Phys.~Rev.~D {\bf 11}, 
3105 (1975).

\bibitem{antreasyan_79}~D.~Antreasyan {\it et al.},~Phys.~Rev.
~D {\bf 19}, 764 (1979).


\bibitem{levai_99}~G.~Papp,~P.~Levai and~G.~Fai,
~Phys.~Rev.~C {\bf 21}, 021902(R) (1999).

\bibitem{wang_00}~X.~-N.~Wang,~Phys.~Rev.~C {\bf 61}, 064910 (2000).

\bibitem{vitev_02}~I.~Vitev and~M.~Gyulassy,
~Phys.~Rev.~Lett.~{\bf 89}, 252301 (2002).

\bibitem{accardi_04}~A.~Accardi and~M.~Gyulassy,
~Phys.~Lett. B {\bf 586}, 244 (2004).
 
\bibitem{ramona_04}~R.~Vogt,~Phys.~Rev.~C {\bf 70}, 064902 (2004).

\bibitem{Armesto:2006ph}
  N.~Armesto,
  hep-ph/0604108 (submitted to J. Phys. G).

\bibitem{dima_03}~D.~Kharzeev,~E.~Levin, and ~L.~McLerran,
Phys.~Lett. B {\bf 561}, 93 (2003).

\bibitem{hwa_05}~R.~C.~Hwa,~C.~B.~Yang, and ~R.~J.~Fries,
~Phys.~Rev.~C {\bf 71}, 024902 (2005).

\bibitem{miklos_plb98}~M.~Gyulassy and~P.~Levai, 
Phys. Lett. B {\bf 442}, 1 (1998).

\bibitem{ampt_lin03}~Z.~-W.~Lin and~C.~M. Ko,
Phys.~Rev.~C {\bf 68}, 054904 (2003).

\bibitem{Werner:2005jf}
  K.~Werner, F.~M.~Liu,~and~T.~Pierog,
  ~Phys.~Rev.~C {\bf 74}, 044902 (2006).







\bibitem{estienne_05}~STAR~Collaboration,~M.~Estienne {\it et al.},
J. Phys. G {\bf 31}, S873 (2005).

\bibitem{nuxu_sqm06}~STAR~Collaboration, N. Xu {\it et al.},
presented at SQM2006, 9th International Conference on 
Strangeness in Quark Matter, UCLA, Los Angeles (26-30 march, 2006)
(to be published in J. Phys. G).


\bibitem{star04_asym}~STAR~Collaboration,~J.~Adams {\it et al.},
Phys.~Rev.\\
~C~{\bf 70}, 064907 (2004).

\bibitem{xnwang_asym03} X.~N.~Wang,~Phys.~Lett.~B {\bf 565}, 116 (2003). 

\bibitem{rene_sqm04} STAR Collaboration, R.~Bellwied {\it et al.},
~J.~Phys.~G {\bf 31}, S675 (2005).

\bibitem{mironov_sqm04} STAR Collaboration, C. Mironov {\it et al.},
~J.~Phys.~G {\bf 31}, S1195 (2005).





\end{thebibliography}
\end{document}